\documentclass[pdflatex,sn-mathphys-num]{sn-jnl}

\usepackage{graphicx}%
\usepackage{multirow}%
\usepackage{amsmath,amssymb,amsfonts}%
\usepackage{amsthm}%
\usepackage{mathrsfs}%
\usepackage[title]{appendix}%
\usepackage{xcolor}%
\usepackage{textcomp}%
\usepackage{manyfoot}%
\usepackage{booktabs}%
\usepackage{algorithm}%
\usepackage{algorithmicx}%
\usepackage{algpseudocode}%
\usepackage{listings}%
\usepackage{comment}
\usepackage{graphicx}
\usepackage{caption}
\usepackage{subcaption}

\theoremstyle{thmstyleone}%
\theoremstyle{thmstyletwo}%

\theoremstyle{thmstylethree}%

\begin{document}

\title[Active control of tokamak plasma turbulence through transient fueling]{Active control of tokamak plasma turbulence through transient fueling}

\title[Gyrokinetic simulation of the effect of transient fueling on plasma turbulence
in ADITYA-U tokamak]{Gyrokinetic simulation of the effect of transient fueling on plasma
turbulence in ADITYA-U tokamak}


\author[1]{\fnm{Jaya Kumar} \sur{Alageshan}}

\author[2,3]{\fnm{Suman} \sur{Dolui}}

\author[2,3]{\fnm{Joydeep} \sur{Ghosh}}
\author[2,3]{\fnm{Kishore} \sur{Mishra}}
\author[2,3]{\fnm{Sarveshwar} \sur{Sharma}}
\author[2,3]{\fnm{Abhijit} \sur{Sen}}

\author[4]{\fnm{Manjunatha} \sur{Valmiki}}
\author[4]{\fnm{Sandeep} \sur{Agrawal}}
\author[4]{\fnm{Sanjay} \sur{Wandhekar}}
\author[5]{\fnm{Zhihong} \sur{Lin}}

\author*[1]{\fnm{Animesh} \sur{Kuley}}\email{akuley@iisc.ac.in}

\affil[1]{\orgdiv{Department of Physics}, \orgname{Indian Institute of Science} \orgaddress{\city{Bangalore}, \postcode{560012}, \country{India}}}

\affil[2]{\orgname{Institute for Plasma Research}, \orgaddress{\street{Bhat}, \city{Gandhinagar}, \postcode{382428}, \country{India}}}

\affil[3]{\orgname{Homi Bhabha National Institute}, \orgaddress{\street{Anushaktinagar}, \city{Mumbai}, \postcode{400094}, \country{India}}}

\affil[4]{\orgname{Centre for Development of Advanced Computing}, \orgaddress{\street{}, \city{Pune}, \postcode{411005} \country{India}}}
\affil[5]{Department of Physics and Astronomy, University of California Irvine, CA 92697, USA}


\abstract{

The gradient-driven microturbulence in ADITYA-U tokamak plasmas has been suppressed by injecting short gas puffs. The suppression of microturbulence increases the core temperature and subsequently the energy confinement time following the gas puff. The gas injection modifies the radial density profile, making it relatively flatter near the mid-radius. Global electrostatic gyrokinetic simulations show that this modification to the radial density profile due to gas injection suppresses the existing trapped electron mode (TEM). Simulation results show that the TEM-dominated turbulence suppression reduces the turbulence-driven heat transport, leading to an increase in core temperature. Applying multiple periodic gas-puffs leads to multiple periodic events of TEM suppression, improving the overall energy confinement time, and is used as an active control mechanism to influence microturbulence in ADITYA-U tokamak.

}



\maketitle

\section*{Introduction}

Control of plasma turbulence remains a central challenge for achieving
enhanced energy confinement in
tokamaks~\cite{TynanFujisawaMcKee2009,Fujisawa2021}, where
gradient-driven microturbulence
produces anomalous particle and heat transport that limits performance.
Traditional approaches in both tokamaks and stellarators rely on impurity
injection to modify turbulence through changes in collisionality,
dilution, and instability drive~\cite{Nespoli2021ReducedTurbulenceLHD,Singh_2024,Singh2024ImpurityGyrokinetic},
while edge modification
techniques—including on and off-axis auxiliary heating and current
drive~\cite{Nakamura2024AdvancedTokamak}—have long been explored for transport mitigation. Among these,
gas puffing offers
a uniquely versatile actuator~\cite{Xiao2022SMBIReview}: short neutral gas pulses injected at the
periphery enable temporally and spatially localized perturbations that
preserves core purity while providing a controlled handle on turbulence-driven
transport, rendering gas-puff–based approaches a fundamentally different and
largely unexplored route for studying causal transport dynamics and active
plasma control in fusion devices.

Experimental and gyrokinetic studies demonstrate that core and edge profile
modifications directly impact turbulence
~\cite{Singh2024ImpurityGyrokinetic,Rhodes2011LmodeValidation,Stechow2020ProfileShapingW7X,Stimmel2022ArgonEDA}.
Edge density gradients
modified by gas puffing suppress edge fluctuations and particle flux, while
core density flattening reduces the drive of trapped-electron-mode (TEM) and
electron temperature gradient (ETG) instabilities by lowering gradient scale
lengths. Recent research suggests that for steep
density profiles, the TEM driven by density gradient or coexisting ITG and
TEM turbulence are dominant, resulting in outward particle diffusion and edge profile modification \cite{Li_2022,Schuster_2023}.

We propose a general receipe that controlled flattening of core density
profiles via short gas puffing can suppress micro-instabilities
such as trapped-electron-mode (TEM) turbulence, thereby reducing electron heat
transport and enhancing core confinement. In regions of significant
trapped-electron fraction, reduced density gradients weakens the drive of
these modes even when temperature gradients remain finite, shifting the
turbulence spectrum toward more benign ion-scale-dominated regimes. As a
case study, we analyze ADITYA-U experiments where short gas puffs produced
core density flattening and enhanced sawtooth periods~\cite{Dolui2025}.
Gyrokinetic simulations using the GTC code~\cite{Singh_2023,Singh_2024,Tiwari_2025,Singh_2025} reveal
that the before gas puffing case
exhibits strong TEM-driven turbulence, while after gas puffing, the
flattened core density substantially reduces TEM drive shifts dominant
mode structures outward (propagates lesser to the core and not enough
drive in the core), and lowers net turbulent transport levels—directly
correlating with the observed core temperature enhancement and improved
confinement.


Overall, gas puffing experiments and simulations have emerged as an essential component
of plasma control and diagnostic strategies in magnetic confinement fusion 
research~\cite{Zweben2017GPIReview}. The
continued integration of high-resolution diagnostics with advanced plasma--neutral modeling
is expected to further clarify the role of gas puffing in regulating turbulence, transport,
and stability in tokamak.

\section*{Results}
\label{Sec:Analysis}

\begin{figure}[!ht]    
    \hspace*{-1cm}
    \centering 
    \includegraphics[width=1.2\linewidth]{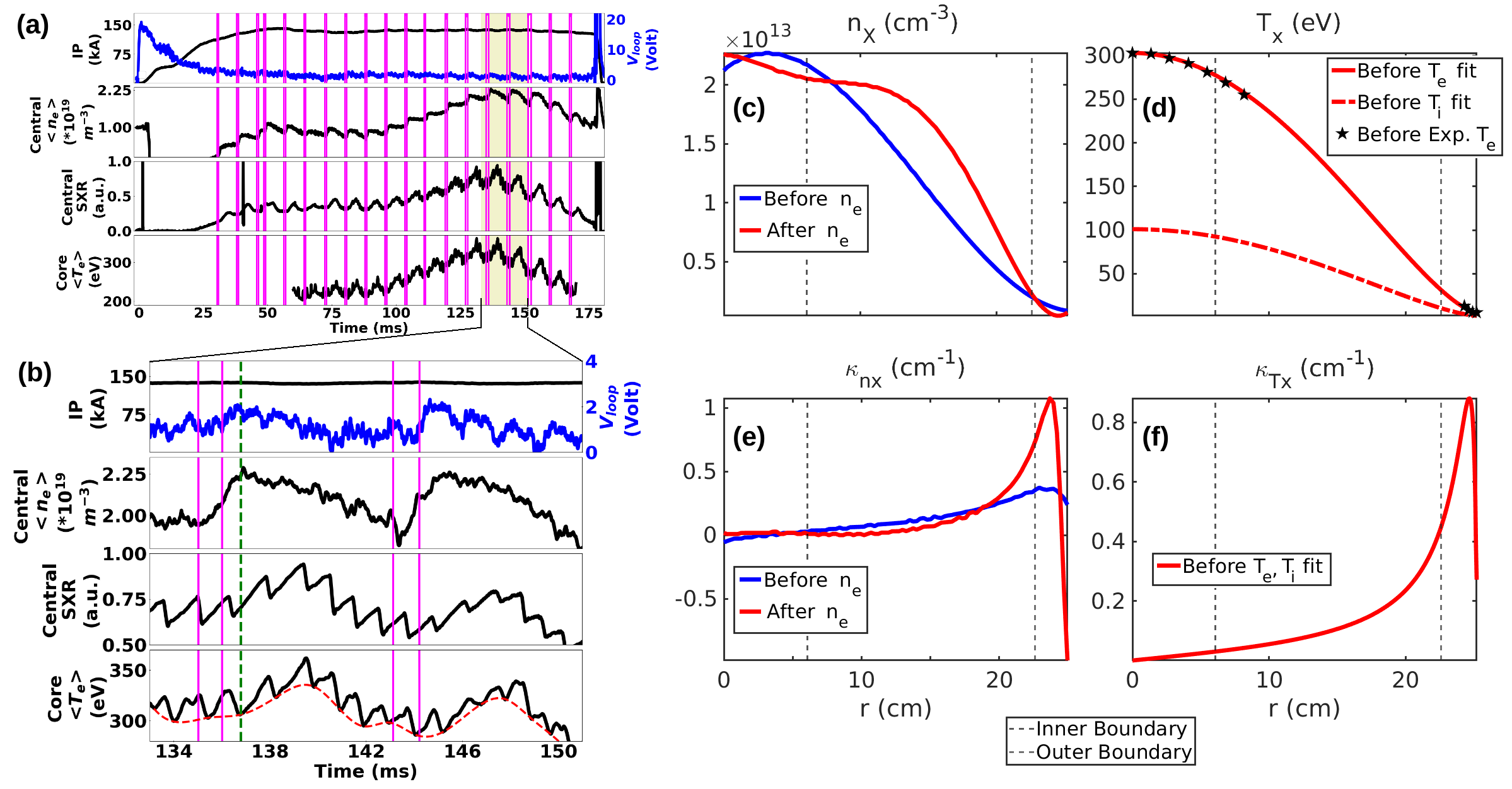}
    \caption{
    (a) Temporal evolution of plasma parameters (Shot $\#36136$); 
    $I_P$ (black), loop voltage $V_{loop}$ (blue), central chord
    averaged electron density $\langle n_e\rangle$; central chord
    averaged SXR intensity and core temperature $\langle T_e \rangle$.
    (b) The time span $133-151$ ms during the plasma current flat-top 
    (yellow shaded area in (a)) is expanded for all the parameters.
    The experimental plasma density profiles are shown in (c) and 
    the electron temperature data with a fit is given in (d);
    here $T_i$ profile is assumed to be one third of $T_e$;
    The gradients are given in (e) and (f), where 
    $\kappa_X := - d(\log X)/dr$. In (c), the core density is flattened
    after gas puffing. Correspondingly there is a dip in the gradients
    shown in (e). The region between the vertical dashed lines indicate
    the simulation domain, with inner boundary at $\psi/\psi_X=0.1$ and
    outer boundary at $\psi/\psi_X=0.95$.}
    \label{Fig:profile}
\end{figure}

{\large \bf Experimental observations:}
Figure~\ref{Fig:profile}(a) shows the temporal evolution of plasma current
$I_P$, loop-voltage $V_{loop}$, central chord-averaged SXR emission
intensity, density  and temperature with the application of periodic
hydrogen gas puffing of ADITYA-U tokamak.

To explore the alterations of these plasma parameters due to a gas-puff,
the temporal variations current flat-top phase ($133 – 155$ ms), are
illustrated in Fig.~\ref{Fig:profile}(b) containing two gas-puff pulses.
Note that the injected gas amount is controlled by the voltage pulse
width applied to the piezo valve; a pulse width of $\sim 1$ ms injects 
$10^{17}$ molecules. Following a gas-puff, the central line-averaged
density ($\bar{n}_e$) rises rapidly and reaches its maximum in $\sim 1$
ms. After the density reaches to its peak value, the central
chord-averaged  electron temperature ($T_e$) starts increasing,
indicated by the green dashed line in Fig.~\ref{Fig:profile}(b),
and peaks subsequently in $3 – 4$ ms after the gas injection. These
observations suggest that the core temperature increases with
gas-puff in ADITYA-U tokamak. Furthermore, it has been observed
that following the gas-injection, the increase in density is
primarily located in the mid-radius region of the plasma column
($0.3<\rho<0.8)$, causing the radial density profile flatter in
this region compared to those without the gas-puff. The density
remains almost constant in the plasma core ($\rho<0.3$) after
the gas-puff. Details of these experiment results can be found
in references~\cite{SUM1,SUM2,SUM3}.

\bigskip

\noindent
{\large \bf Modelling and Simulations:}
The rise in core-temperature following a gas-puff occurs in a
time-scale ($\sim 3-4$ ms) shorter than the energy confinement
time ($\sim 5–10$ ms) in typical discharges of ADITYA-U tokamak.
Similar observations are reported from several tokamaks with
impurity injection~\cite{SUM4,SUM5}. The physical phenomena behind this fast
temperature rise with impurity injection are attributed to turbulence
spreading~\cite{SUM6} or to fluctuation suppression~\cite{SUM4}. Turbulence is
spreading is not observed in ADITYA-U as the core density fluctuations
are observed to be decreasing after the gas-puff~\cite{SUM3}. Therefore to
investigate the role of fluctuation suppression in core temperature
rise with gas-puff in ADITYA-U, Gyrokinetic simulations~\cite{Singh_2023} are 
performed incorporating the equilibrium and radial profiles of density
and temperature of the above-mentioned discharges of ADITYA-U. The
radial density profile is measured using microwave interferometry~\cite{Dolui2025}.
The radial profile of the electron temperature is reconstructed by
combining temperature data from multiple soft X-ray (SXR) chords~\cite{Dolui2025}
covering the core region ($\rho=0-0.35$) with measurements from triple
Langmuir probes located in the edge region ($\rho=0.97-1.008$)~\cite{SUM1}.

\bigskip

\noindent
{\large \bf Driving micro-instability:}
Gradient-driven global gyrokinetic simulations of microturbulence for the
ADITYA-U tokamak with gas injection, using the GTC code, employ the
realistic equilibrium obtained using IPREQ~\cite{SHARMA2020111933} and plasma profiles in
Fig.~\ref{Fig:profile}(a)-(d), to model conditions both before and after
gas injection. Both passing and trapped electrons are treated kinetically.
The simulation time step is set to 
$0.005R_0/C_s$, where $C_s=\sqrt{T_e/m_i}$ is the ion sound speed and $R_0$ the major radius.
Electron subcycling is applied with two substeps per ion step, and 50 marker particles are
loaded per cell. The radial simulation domain spans $r/a \in [0.2428,0.9028]$, corresponding to
flux coordinates $\psi/\psi_X \in [0.1,0.95]$, where $a$ is the minor radius and $\psi_X$ the
magnetic flux at the last closed flux surface. The innermost region $\psi/\psi_X \in [0,0.1]$
is excluded, since the plasma profile is nearly flat there (see 
Fig.~\ref{Fig:profile}(c)-(d)), yielding negligible gradients.
Non-uniform plasma density and temperature profiles drive the micro-instabilities.
To analyze such instabilities, we carry out a converged GTC simulations with $120$
flux surfaces, $4000$ poloidal grid points, $32$ grid points in the parallel
direction and $50$ particles-per-cell. 
\begin{figure}[!ht]
    \centering
    \includegraphics[width=\linewidth]{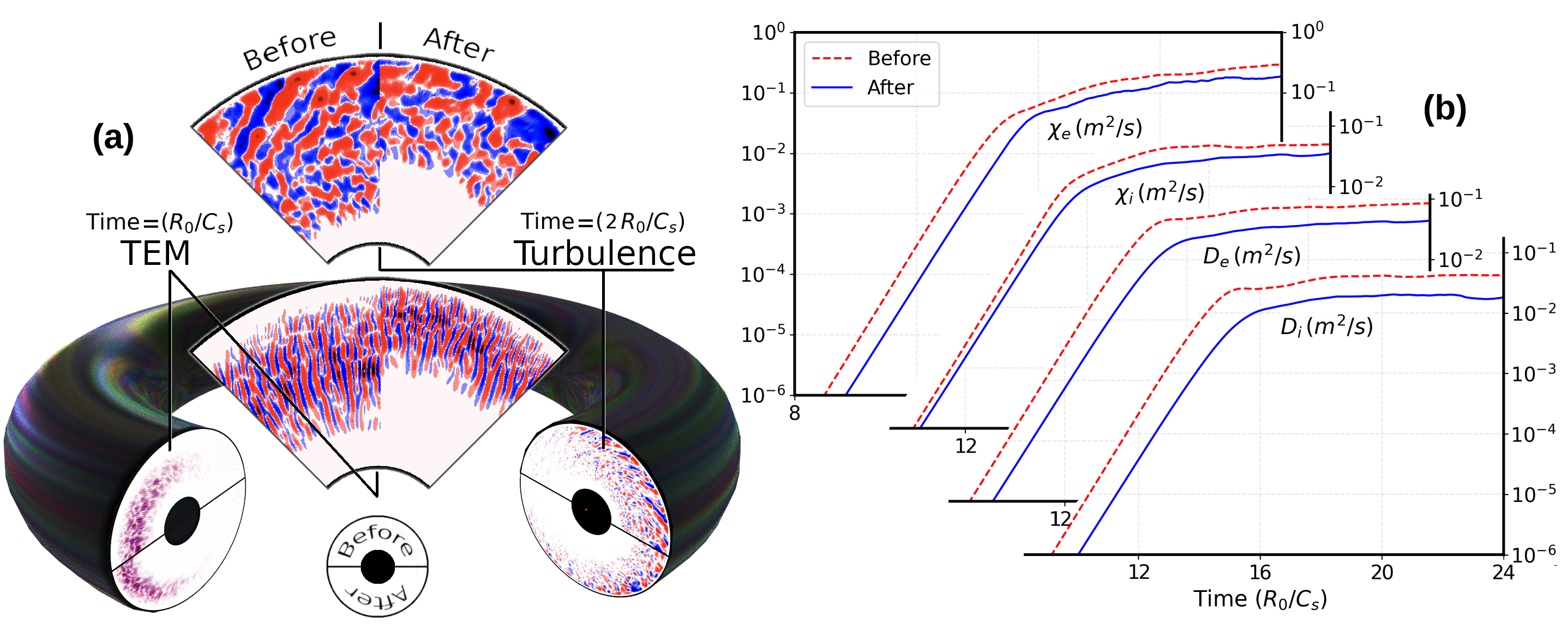}
    \caption{(a) The poloidal cross-section of ADITYA-U flux surface and the
    corresponding mode structure in the electric potentials are shown for both
    before and after cases in the linear (TEM) and turbulent regimes. The electrostatic potential is normalized by $T_e/e$; 
    (b) Comparing the energy transport and particle
    transport for the Before- and After-cases we find all diffusivities
    have similar growth rates. In particular, The saturation
    level for After case is reduced by $\approx 84\%$ in $D_e$ 
    and $\approx 94\%$ in $D_i$. The saturation values $\chi_e$ is an
    order of magnitude higher than $\chi_i$.}
    \label{Fig:Mode_Structure}
\end{figure}

The linear-regime mode structure in ADITYA-U's electric potential, $\delta\phi$ for the
Before- and After-gas puff cases are shown in Fig.~\ref{Fig:Mode_Structure},
at $t=12\:R_0/C_s$. The analysis of the microinstabilities
in both cases show that the TEM (Trapped Electron Mode) is the dominant
instability, compared to the ITG (Ion Temperature Gradient), as signaled by the
sign of diamagnetic frequency. The right panel of Fig.~\ref{Fig:Mode_Structure}
show that the gas puffing leads to small reduction in the growth rates of the
instabilities from $\gamma \: C_s/R_0 \approx 0.82$ (Before) to $0.79$ (After).
The location of the flux surface averaged mode
rms peak shifts from $\psi/\psi_X \approx 0.5$ (Before) to $0.75$ (After) and
comparing the radial extent of $\phi$ for before and after cases show
localization of modes such that after gas puffing the TEM is expelled from
the core. The mode structure sizes show that mode number reduces from
$m\approx 160$ to $120$ after gas puffing.

\bigskip

\noindent
{\large \bf Turbulent transport:}
The kinetic electron based gyrokinetic simulations of the before gas
puffing case indicate strong TEM-driven turbulence. After gas puffing, the
flattened density in the core lead to reduced TEM drive and reduced
trapped-electron activity, shifting the net turbulence transports to
lower levels compared to before case.

\begin{figure}[!ht]
    \centering
    \includegraphics[width=\linewidth]{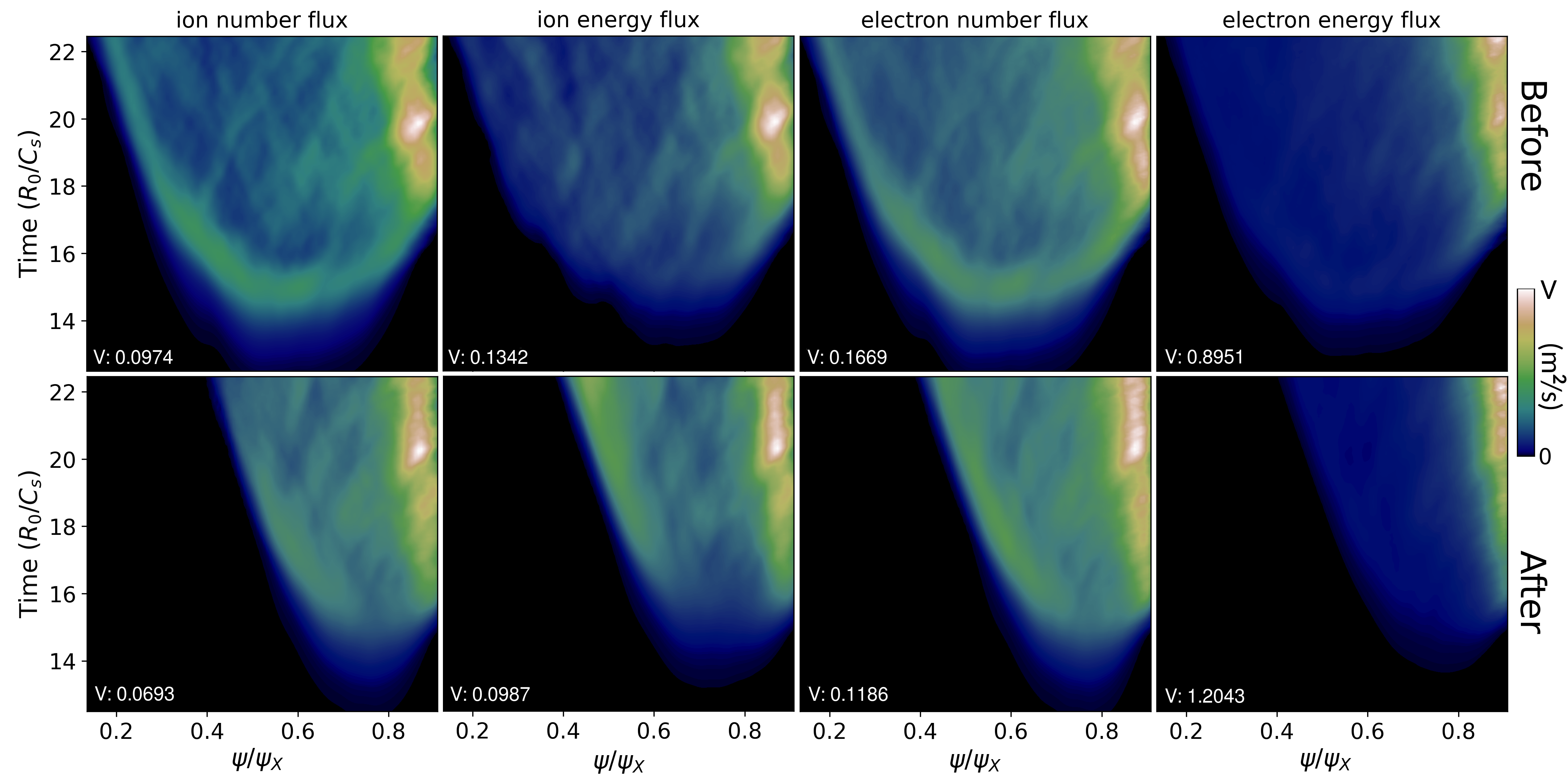}
    \caption{Evolution plots of the radial profiles of turbulent transport fluxes 
    show that in general the radial spread 
    are lower in the after case compared to the before case. The fluctuations
    start to grow at location where the TEM is strong and spreads radially.
    The inward propagation of the fluctuations are curtailed after gas puffing.
    }
    \label{fig:2D}
\end{figure}

In the nonlinear turbulent regime, the fluctuating $\mathbf{E}\times\mathbf{B}$
drift leads to cross-field transport of both energy and particles. The
turbulent transport fluxes are computed from gyrokinetic moments of the fluctuating
distribution and potential.
The radial particle flux for species $s$ is given by
$\Gamma_s = \left\langle \int \delta f_s \, \left( \mathbf{v}_E \cdot 
\nabla \psi \right)\, d^3v_s \right\rangle_\psi$,
where $\mathbf{v}_E = (c/B) \mathbf{b} \times \nabla \phi$ is the 
$\mathbf{E}\times\mathbf{B}$ drift velocity, and $\langle \cdot 
\rangle_\psi$ denotes flux-surface averaging.
The corresponding transport coefficients are then expressed as
$D_s = \frac{\Gamma_s}{\partial n_s / \partial r}$,
and 
$\chi_{s} = \frac{\langle \vec{\mathbf{q}}_s\cdot\nabla\psi \rangle_\psi} {n_s  
        \langle\left| \nabla \psi \right|^2 \rangle_\psi  \frac{\partial T_s}
        {\partial \psi}}$.
The plots in Fig.~\ref{Fig:Mode_Structure}(b) show that the transport
coefficients are lower after gas-puffing. Furthermore, Fig.~\ref{fig:2D}
reveals that the propagation of turbulence across the flux surfaces and
into the core region is suppressed after the gas-puffing.

\section*{Conclusions}
\label{Sec:Conclusion}

Short gas puffing experiments in the ADITYA-U tokamak reveal a clear and reproducible increase in core electron density and temperature leading to an improvement in plasma confinement. The nonlinear GTC simulations show that trapped electron mode (TEM) turbulence is strongly suppressed post-injection, due to the flattened density profile in the mid-radius region of plasma column. The suppression of TEM leads to lower heat diffusivity, allowing core temperature to rise. Spatially, the strongest transport modification is observed in the mid-radius region rather than near the separatrix, suggesting that gas puffing influences core turbulence indirectly through profile relaxation mechanism. The simulations thus provide a self-consistent picture of turbulent transport in ADITYA-U. The simulations capture both the microinstability dynamics and their nonlinear saturation, enabling quantitative evaluation of turbulent particle and energy fluxes. The insights gained from such simulations are essential for understanding transport regulation mechanisms and for developing predictive models for tokamak confinement optimization. The experimental and simulation results clearly demonstrate that neutral fueling can function as an active turbulence control mechanism rather than merely a particle source.

\section*{Acknowledgements}
\noindent This work is supported by the Board of Research in Nuclear Sciences (BRNS Sanctioned no. and 57/14/04/2022-BRNS), Science and Engineering Research Board EMEQ program (SERB sanctioned no. EEQ/2022/000144), National Supercomputing Mission (NSM). We acknowledge National Supercomputing Mission (NSM) for providing computing resources of `PARAM PRAVEGA' at S.E.R.C. Building, IISc Main Campus, Bangalore, which is implemented by C-DAC and supported by the Ministry of Electronics and Information Technology (MeitY) and Department of Science and Technology (DST), Government of India, and  ANTYA cluster at Institute of Plasma Research, Gujarat.

\bibliography{sn-bibliography}

\end{document}